\documentclass[prd,twocolumn,floatfix,showkeys]{revtex4}
\usepackage{hyperref}
\usepackage{epsfig}
\usepackage{amsmath}
\begin{document}                
\pagestyle{myheadings} \markright{Unification of classical and
quantum mechanics}
\title{Unification of classical and quantum mechanics}
\author{Jos\'e B. Almeida}
\affiliation{Universidade do Minho, Departamento de F\'isica,
4710-057 Braga, Portugal.} \email{bda@fisica.uminho.pt}

\keywords{Unified theories; quantum mechanics; general relativity}

\date{November 12, 2002}
\begin{abstract}                
This paper is a serious attempt at reconciling quantum and
classical mechanics through the concept of dynamic space and the
acceptance of non-zero Ricci tensor for vacuum. Starting with
scalar particles, the paper shows that those two points allow
predictions of General Relativity to be made with an equation
which includes Kein-Gordon as a special case; this equation is
designated the \emph{source equation}. The paper then moves on to
show that Dirac equation is compatible with the source equation
written in a more general form. 4-dimensional optics is
introduced as an alternative to space-time, which is shown to
allow similar predictions in the case of gravity, but has the
great advantage of ascribing also electro dynamics to space
curvature.

\end{abstract}
\maketitle
\section{Introduction}
General Relativity and Quantum Mechanics are the two most
successful theories in Physics; each in its own field of
application seems to be the ultimate theory, yet every effort to
merge them together has encountered unsurmountable difficulties.
Particle physics is reasonably explained by the Standard Model, in
spite of this theory's uncomfortable number of parameters. It has
frequently been said that the unification of General Relativity
with Quantum Mechanics and the Standard Model can only arise from
a new paradigm which drastically departs from some of the present
day assumptions setting Physics in a new framework. Some people
believe that String Theory provides this new way of thinking and
is the road to unification \cite{Greene99}. In spite of some
success we feel that String Theory is not yet the answer
everybody is waiting for, not the least because it is an
extremely complex theory while intuitively one  prefers theories
that are simple, at least in their basic principles. Inevitably
mathematics will always become more involved as  complex problems
are attacked. This paper makes a contribution for a unified
theory based on simple equations but it does use concepts that
are unusual and may be difficult to accept due to long
established habits. The paper introduces a generalization of
Dirac's equation such that not only electrodynamics but also
gravitation can be accounted for. Full compatibility with the
Standard Model is not explored in this work but the path that
must be followed is delineated.

The author has previously made some attempts at formulating a
unified theory of physics, namely through the use of the
\emph{4-Dimensional Optics} (4DO) concept \cite{Almeida02:1,
Almeida02:2}; the present work presents a more fundamental
approach, departing from a mathematical equation deprived of
physical interpretation. The concept of space is used as a purely
geometric concept and the notions of distance, space or mass are
introduced in connection to solutions of the equation that were
found in purely mathematical grounds.

In this paper we use the concept of dynamic space which is
entirely linked to some dynamics under study and not preexisting.
Dynamic space exists for the dynamics and the latter occupies the
former completely. Ultimately all dynamics are part of the grand
Universe dynamics and this supports a universal space; when we
choose to consider some dynamics isolated from the Universe we
also define a dynamic space which contains and is contained by the
dynamics. Consider as examples a planetary system or a galaxy; in
either case the study of their dynamics in isolation involves the
definition of a particular dynamic space which does not extend
outside the edges of each particular system. The observation of
some dynamics by a non-interfering observer is impossible; the
observer has to participate in the dynamics in such a way that
dynamic space is extended to include him. We will find ways to
deal with non-interfering observers by a suitable change of
coordinates between dynamic space and observer space; this is
seen as a useful approximation, frequently more manageable than
the exact solution which would involve including the observer as
one element of the dynamics. A useful reading on the subject of
space creation is provided by Ref.\ \cite{Gouin}.

Our strategy for this work involves showing that Klein-Gordon
equation and gravitation can be made compatible through the
dynamic space concept associated with a rejection of \emph{empty
space} and consequent zeroing of the Ricci tensor for vacuum. We
will then show that the same assumptions can be applied to Dirac
equation and propose its generalization such that electrodynamics
and gravitation become effectively integrated. 4-dimensional
optics with its associated space appear as a consequence of this
process, which could not be achieved with conventional
space-time.  We will also suggest that the source equation may
have some solutions which explain the existence of elementary
particles as dynamic systems in their own right.
\section{Classical mechanics and Klein-Gordon equation}
In this section we examine how Kein-Gordon equation and gravity
can be merged into a single equation, designated the \emph{source
equation}; later on we will show this to be a consequence of a
more fundamental equation including Dirac's as a special case.
The source equation is a 4-dimensional differential equation with
respect to the four variables $x^\mu$, which should be seen as
pure numbers without any physical meaning; the concepts of
distance or time are absent at this stage and are later defined
within the theory. The simplest form of the source equation is
\begin{equation}
    \label{eq:founding}
    g^{\mu \nu}
    \nabla_{\mu \nu} \psi = -\psi,
\end{equation}
where $g^{\mu \nu}$ is a metric tensor and $\psi$ is a function
of the 4 variables with some general characteristics that will be
defined below. Einstein's summation convention is used in the
equation; the notation $\nabla_\mu$ means  covariant derivative
with respect to the variable $x^\mu$. Throughout this note greek
indices and superscripts take values between 0 and 3 and roman
indices run from 1 to 3. For the moment the elements $g^{\mu
\nu}$ are treated as real and the function $\psi$ as complex,
although further along the equation will have to be generalized
to accept other entities; in every situation $g^{\mu \nu}$
incorporates some form of \emph{feedback} from the function
$\psi$, characterizing what we will designate by \emph{dynamic
space}. Equation (\ref{eq:founding}) can be associated with the
Lagrangian density
\begin{equation}
    \label{eq:lagdens}
    \mathcal{L} = \frac{1}{2} \left(g^{\mu \nu} \partial_\mu
    \psi \partial_\nu \psi- \psi^2 \right).
\end{equation}
In a large number of situations we will not have to distinguish
between covariant and partial derivatives, but the cases we
analyze at the end of the paper make this distinction necessary.

We are not in a position to give a general solution to this
equation but will instead discuss various particular solutions in
order to understand how it becomes fundamental in generating the
basic laws of physics. As a starting point we discuss the
solution $\psi = \exp(\mathrm{i}p_\alpha x^\alpha)$. Notice that
if $g^{\mu \nu}$ is replaced by the metric $\eta^{\mu \nu}/m^2$,
with $\eta^{\mu \nu}=\mathrm{diag}(1,-1,-1,-1)$ the Minkowski
metric, the equation is equivalent to the Klein-Gordon equation,
so that we need only to verify that gravitation is also
comprehended. The solutions of Klein-Gordon equation include $m$
in the exponent of $\psi$ so that there feedback from the solution
of the equation to the space metric, a characteristic of dynamic
space. We will look at a more general situation, assuming that the
metric has the form
\begin{equation}
    \label{eq:gravmetric}
    g^{\mu \nu} = \frac{1}{m^2}\, \mathrm{diag}
    \left(\frac{1}{n^2},-1,-1,-1 \right),
\end{equation}
where $m$ is a constant and $n$ is a function of the coordinates
$1$, $2$ and $3$.

The list of non-zero Christoffel symbols is
\begin{equation}
    \label{eq:christoffel}
    \begin{aligned}
    {\Gamma^0}_{0 i} &= {\Gamma^0}_{i 0}
    = \partial_i \log n  \\
    {\Gamma^i}_{0 i} &= n^2 \partial_i n.
    \end{aligned}
\end{equation}
Performing the derivations in Eq.\ (\ref{eq:founding}) and
assuming that $n$ is a slowly varying function, so that the
Christoffel symbols are small compared to the derivatives of
$\psi$,
\begin{equation}
    \label{eq:mgrav}
    \frac{1}{m^2}\left[
    \frac{1}{n^2}\, \left(p_0 \right)^2 - \delta^{i j}
     p_i p_j \right] = 1.
\end{equation}

The $p_\mu$ in Eq.\ (\ref{eq:mgrav}) are the conjugate momenta in
a space whose metric is $g^{\mu \nu}$; we call this space the
dynamic space. This designation is entirely appropriate since the
space metric includes the factor $1/m^2$, which will be
associated with a particular dynamics. The concept of dynamic
space will become clearer further along, when we make the
transition to the more conventional observer space.

The relation between the conjugate momenta and the coordinates is
given as usual by \cite{Martin88}
\begin{equation}
    \label{eq:momenta}
    p_\mu = g_{\mu \nu} x^\nu.
\end{equation}
Considering Eq.\ (\ref{eq:gravmetric}) it is
\begin{equation}
    \label{eq:mgravmetr2}
    g_{\mu \nu} = m^2 \mathrm{diag}\left(n^2,-1, -1 , -1 \right),
\end{equation}
which we insert into Eq.\ (\ref{eq:momenta})
\begin{equation}
    \label{eq:momenta2}
    p_0 = m^2n^2 \frac{\mathrm{d}x^0}{\mathrm{d} s},~~~~
    p_i = - m^2    \frac{\mathrm{d}x^i}{\mathrm{d} s} ,
\end{equation}
with the line element $\mathrm{d}s$ defined by
\begin{equation}
    \label{eq:timeelement}
    \left(\mathrm{d} s \right)^2 = m^2 \left[n^2 \left(\mathrm{d}
    x^0 \right)^2 - \delta_{i j} \mathrm{d}x^i \mathrm{d}x^j \right].
\end{equation}

We stress that all the elements in the equations above are pure
numbers to which we have not attributed any physical meaning. All
the operations done so far are purely mathematical, including the
concept of space; there are no dimensions of any sort attributed
to the space characterized by the the metric just defined.

The geodesic equations of any space can be found by defining a
Lagrangian with constant value $1/2$
\begin{equation}
    \label{eq:lagrangian}
    2 L =1 = g_{\mu \nu} \frac{\mathrm{d}x^\mu}{\mathrm{d} s}
     \frac{\mathrm{d}x^\nu}{\mathrm{d} s},
\end{equation}
and writing the associated Euler-Lagrange equations
\begin{equation}
    \label{eq:euler}
    \frac{\mathrm{d}p_\alpha}{\mathrm{d} s} = \frac{1}{2}\,\partial_\alpha
    \left(g_{\mu \nu} \frac{\mathrm{d}x^\mu}{\mathrm{d} s}
     \frac{\mathrm{d}x^\nu}{\mathrm{d} s} \right).
\end{equation}
Applying to dynamic space we get its geodesic equations as
\begin{equation}
\begin{split}
    \label{eq:geodesic1}
    p_0 = m^2 n^2  \frac{\mathrm{d}x^0}{\mathrm{d} s} =
    \mathrm{constant};\\
    \frac{\mathrm{d}p_i}{\mathrm{d} s}
    = m^2 n \partial_i n \left(\frac{\mathrm{d}x^0}{\mathrm{d} s} \right)^2.
\end{split}
\end{equation}
Since $ d/\mathrm{d} s =\mathrm{d}/ \mathrm{d} x^0 \times
\mathrm{d}x^0/\mathrm{d} s  $ we can replace the second relation
by
\begin{equation}
    \label{eq:geodesic2}
    \frac{\mathrm{d}}{\mathrm{d}x^0}\left(\frac{1}{n^2}\,
    \frac{\mathrm{d} x^i}{\mathrm{d} x^0}\right)
    = -\frac{\partial_i n}{n}.
\end{equation}

There is not much more we can do to the geodesic equation without
particularizing the form of $n$; we will do that in the following
paragraphs.

We will now consider the situation where $n$ is symmetric in all
three $x^i$ coordinates; this will be done by making $n$ a
function of $\sqrt{(x^1)^2 + (x^2)^2 + (x^3)^2}$. We will also
perform a change of coordinates to remove $m$ from the metric
\begin{align}
    \label{eq:t}
    t &= m x^0; \\
    \label{eq:r}
    r &= m \sqrt{(x^1)^2 + (x^2)^2 + (x^3)^2}; \\
    \label{eq:theta}
    \theta &= \arccos \left[x^3/\sqrt{(x^1)^2 + (x^2)^2 + (x^3)^2}
    \right]; \\
    \label{eq:phi}
    \varphi &= \arccos \left[x^1/\sqrt{(x^2)^2 + (x^3)^2} \right].
\end{align}

We are dealing here with a one body problem; we anticipate that
$m$ is somehow linked to inertial mass. In a more general
situation, when multiple bodies participate in the dynamics, the
scaling parameter $m$ will vary according to which body is being
considered. The use of $m$ as scaling parameter in Eq.\
(\ref{eq:r}) means that points of the new space, which we will
designate by observer space, can only be mapped to points in
dynamic space if $m \neq 0$; there will be points of observer
space where $m = 0$; these points will necessarily map to
infinity in dynamic space. Anticipating the physical
interpretation of the equations, we will say that points of
observer space where there is no mass necessarily map to infinity
in dynamic space.

Now we can write the Lagrangian in the new coordinates
\begin{equation}
\begin{split}
    \label{eq:newlagrang}
    2 L = &\, n^2 \left(\frac{\mathrm{d}t}{\mathrm{d} s} \right)^2
    - \left(\frac{\mathrm{d}r}{\mathrm{d}s} \right)^2 \\
    &- r^2 \left(\frac{\mathrm{d}\theta}{\mathrm{d} s}\right)^2
    - r^2 \sin^2 \theta \left(\frac{\mathrm{d}\varphi}{\mathrm{d}s}
    \right)^2;
\end{split}
\end{equation}
where $n$ is a function of $r$. It is reasonable to make $\theta =
\pi/2$, so that the Euler-Lagrange equations become after some
manipulation
\begin{gather}
    \label{eq:taual}
    n^2 \frac{\mathrm{d}t}{\mathrm{d} s} = \gamma  ~~
      \rightarrow ~~ \mathrm{constant};\\
    \label{eq:phial}
    \frac{r^2}{n^2} \frac{\mathrm{d}\varphi}{\mathrm{d} t} =  J   ~~
      \rightarrow ~~ \mathrm{constant};\\
    \label{eq:radial}
    \frac{\mathrm{d}}{\mathrm{d}t}\left(\frac{1}{n^2}\,
    \frac{\mathrm{d} r}{\mathrm{d}t}\right)=
    -\frac{1}{n}\frac{\mathrm{d}n}{\mathrm{d} r}\
    + \frac{n^2 J^2}{r^3}.
\end{gather}

As a last step in particularizing an application of the source
equation we will now assume a special form for $n$ given by
\begin{equation}
    \label{eq:gravpot}
    n = \exp \left(\frac{-M}{r} \right),
\end{equation}
with $M$ a constant. Eqs.\ (\ref{eq:taual}) to (\ref{eq:radial})
become
\begin{gather}
    \label{eq:taual2}
    \mathrm{e}^{-2M/r} \frac{\mathrm{d}t}{\mathrm{d} s} = \gamma  ~~
      \rightarrow ~~ \mathrm{constant};\\
    \label{eq:phial2}
    r^2 \mathrm{e}^{2M/r}
     \frac{\mathrm{d}\varphi}{\mathrm{d} t} =  J   ~~
      \rightarrow ~~ \mathrm{constant};\\
    \label{eq:radial2}
    \frac{\mathrm{d}^2 r}{\mathrm{d}t^2} = -\frac{M}
    {r^2 e^{2M/r}}+\frac{2M}{r^2}
    \left(\frac{\mathrm{d}r}{\mathrm{d} t}\right)^2
     - \frac{J^2}{r^3 e^{4M/r}}.
\end{gather}

These equations are still written on pure numbers but we are now
in position to bring in physical meaning and units.
\label{interpretation}Let us start with the concepts of length and
time and assume that we want to associate some length units to the
values of coordinate $r$ and time units to the values of
coordinate $t$; we can do this by introducing arbitrary
dimensional factors such that $r = \mathcal{R} r_d $ and $ t =
\mathcal{T}t_d$, where the subscript $d$ stands for
"dimensional". The ratio $\mathcal{R}/\mathcal{T}$ defines the
physical constant $c$.

A similar procedure is done for the values of $m$ and $M$, such
that $M = \mathcal{M} M_d$. The fraction $M/r$ becomes
$\mathcal{M} M_d / (\mathcal{R} r_d)$ and must be non-dimensional
leading to the definition of the physical constant $G = c^2
\mathcal{M}/\mathcal{R}$. There is one third relation linking the
dimensional factors to physical constants; Planck's constant is
defined as $\hbar = \mathcal{M}\mathcal{R}^2/\mathcal{T}$. The
three physical constants $c$, $G$ and $\hbar$ are an alternative
to the three dimensional factors $\mathcal{R}$, $\mathcal{T}$ and
$\mathcal{M}$, the latter allowing all equations in physics to be
written as non-dimensional.

At this point it is important to link $c$ to the speed of light
in vacuum. Photons have no mass and classically do not
participate in the dynamics; it comes as no surprise that dynamic
space is not applicable to photons, as long as they are excluded
from the dynamics. We can study photon dynamics as a limiting
case where the mass tends to zero. Recalling Eq.\
(\ref{eq:mgrav}), consider the special case where $m=0$ and
$(p_0)^2 =n^2 \delta^{ij}p_i p_j$; the 4-dimensional equation
becomes undetermined but we can write a 3-dimensional counterpart
\begin{equation}
    \label{eq:photons1}
    \frac{n^2 \delta^{i j}}{\left(p_0 \right)^2}\, p_i p_j = 1.
\end{equation}
Proceeding as before to convert into observer space's coordinates
but using $p_0$ instead of $m$ as the scaling parameter we get
\begin{equation}
    \label{eq:photons2}
    \left(\mathrm{d}s'\right)^2 = \frac{1}{n^2}
    \left[\left(\mathrm{d}r\right)^2 + r^2 \left( \mathrm{d}
    \theta \right)^2 + r^2 \sin^2 \theta \left(\mathrm{d} \varphi
    \right)^2 \right].
\end{equation}

Since both members are dimensionless we can choose to express the
first member in terms of time dimensions while the second member
is expressed in distance dimensions; after a little
re-arrangement, considering the definition of $c$,
\begin{equation}
    \label{eq:photons3}
    c^2 n^2  \left(\mathrm{d} t_d \right)^2
    =
    \left(\mathrm{d}r_d\right)^2 + r_d^2 \left( \mathrm{d}
    \theta \right)^2 + r_d^2 \sin^2 \theta \left(\mathrm{d} \varphi
    \right)^2 .
\end{equation}

Equation (\ref{eq:photons3}) defines the worldline of a particle
characterized by $(p_0)^2 = n^2 \delta^{ij}p_i p_j$; notice,
though, that by simultaneously making $m = 0$ Eq.\
(\ref{eq:mgrav}) generates an indetermination and so we can say
that Eq.\ (\ref{eq:photons3}) defines the worldline of a massless
particle. When both members are divided by $(\mathrm{d}t)^2$, the
first member becomes the squared velocity of a massless particle
in the space whose metric is given by Eq.\ (\ref{eq:mgravmetr2}).

Physical interpretation of Eqs.\ (\ref{eq:taual2} to
\ref{eq:radial2}) becomes now possible. The first of these
equations guarantees $\mathrm{d}t/\mathrm{d}s$ is finite and so
$\mathrm{d}r /\mathrm{d}t =
(\mathrm{d}r/\mathrm{d}s)/(\mathrm{d}t/\mathrm{d}s)$ cannot grow
to infinity; hence the velocity remains below the local velocity
of light. The second equation is almost Kepler's second law but
the presence of the exponential in the first member introduces a
small perturbation which induces a precession of the perihelium in
closed orbits; this is the same effect that is known from General
Relativity as can easily be checked against Schwartzschild's
predictions based on the metric \cite{Martin88}
\begin{equation}
\begin{split}
    \label{eq:schwart1}
    \mathrm{d}s^2 = \left(1 - \frac{2M}{r} \right)
    \mathrm{d}t^2 - \left(1 - \frac{2M}{r}
    \right)^{-1} \mathrm{d}r^2 \\
     - r^2 \left(\mathrm{d} \theta^2 +
    \sin^2 \theta \mathrm{d}\varphi^2\right).
\end{split}
\end{equation}
The counterpart of Eq.\ (\ref{eq:phial2}) is
\begin{equation}
    \label{eq:phischwart}
    \left(1-\frac{2M}{r} \right)^{-1} r^2
    \frac{\mathrm{d}\varphi}{\mathrm{d}t} = J;
\end{equation}
since $\mathrm{exp}(2M/r)$ and $(1-2M/r)^{-1}$ both expand to the
first order of $1/r$ as $1+ 2M/r$, the predictions for the
perihelium advance will be similar.

The last equation deserves more detailed analysis. The first term
in the second member provides the compatibility with Newton's
dynamics by making the acceleration inversely proportional to the
second power of the distance between masses $m$ and $M$. The
second term introduces a repulsive acceleration which can become
as important as the Newtonian term when the radial velocity
approaches $c/\sqrt{2}$, thus effectively preventing the speed of
light to be attained. The last term introduces a short distance
correction which manifests itself only if $J \neq 0$, i.e. if
there is angular momentum.

When $r$ is of the same order of magnitude as $M$ or smaller the
present analysis becomes invalid. First of all we must realize
that $r = 2M$ is the horizon of a black hole with Schwartzschild's
metric, so we are now discussing situations where physics is
largely untested. On the other hand, in the derivation of Eq.
(\ref{eq:momenta2}) it was assumed that $n$ was a slowly varying
function, which is not clearly the case for small values of $r$;
in this situation the Christoffel symbols cannot be neglected and
a different dynamics appears, which is no longer described by
Eq.\ (\ref{eq:timeelement}).

One very important point results from the exponential form of $n$
in Eq.\ (\ref{eq:gravpot}) where the Newton potential appears as
exponent of the metric coefficient $g_{00}$; as a consequence
Schwartzschild's condition of zero Ricci tensor in vacuum is not
verified. This may mean that zero point field must be considered
in the construction of the metric for vacuum in the Universe's
dynamic space. In Ref.\ \cite{Almeida02:1} the author shows that
the exponential form is related to the evanescent field of
4-dimensional waveguides; the fact that the cited work uses
pseudo-Euclidean space is of no consequence in this respect, as
we shall see later on. The origin of the exponential dependence
does not need to be considered in the remainder of the present
work if the reader accepts it based on results rather than on
physical explanation.
\section{Dirac equation}
We want to show that the dynamic space concept is entirely
compatible with Dirac equation; the aim is to ensure that quantum
mechanics and dynamic space merge together flawlessly, thus
merging Quantum Mechanics and gravitation.

We will start with the following form of Dirac equation
\cite{Cottingham98}
\begin{equation}
    \label{eq:dirac}
    \left(\mathrm{i}\frac{\partial}{\partial t} +
    \mathrm{i}\, \boldsymbol{\alpha}\cdot\nabla -\alpha^0 m \right)
    \psi =0;
\end{equation}
where $\psi$ is a 4-element spinor and $\alpha^\mu$ are $4 \times
4$ Pauli matrices for which we adopt the Chiral representation
\begin{equation}
    \label{eq:pauli4}
    \alpha^i = \begin{pmatrix}
      - \sigma^i & \mathbf{0} \\
      \mathbf{0} & \sigma^i  \
    \end{pmatrix},~~~~
    \alpha^0 = \begin{pmatrix}
      \mathbf{0} & \sigma^0 \\
      \sigma^0 & \mathbf{0} \
    \end{pmatrix},
\end{equation}
with $\sigma^i$ the $2 \times 2$ Pauli matrices
\begin{equation}
    \label{eq:pauli2}
    \sigma^1 = \begin{pmatrix}
    0 & 1 \\ 1 & 0 \end{pmatrix},~~
    \sigma^2 = \begin{pmatrix}
    0 & -\mathrm{i} \\
    \mathrm{i} & 0 \end{pmatrix},~~
    \sigma^3 = \begin{pmatrix}
    1 & 0 \\
    0 & -1 \end{pmatrix};
\end{equation}
in addition these two other matrices are used
\begin{equation}
    \label{eq:other2}
    \sigma^0 = \begin{pmatrix}
    1 & 0 \\
    0 & 1 \end{pmatrix}, ~~~~
    \mathbf{0} = \begin{pmatrix}
    0 & 0 \\ 0 & 0 \end{pmatrix}.
\end{equation}
The following relations apply
\begin{equation}
    \label{eq:relation1}
    \alpha^\mu \alpha^\nu + \alpha^\nu \alpha^\mu
    = 2 \mathrm{diag}\left(\mathrm{I},\mathrm{I},\mathrm{I},\mathrm{I}
    \right).
\end{equation}
We shall prefer the full tensorial alternative to the above
equation
\begin{equation}
    \label{eq:relation1tens}
    {\alpha_\lambda}^{\beta \mu} {\alpha_\beta}^{\kappa \nu}
    + {\alpha_\lambda}^{\beta \nu} {\alpha_\beta}^{\kappa \mu}= 2
    \delta_\lambda^\kappa \delta^{\mu \nu}.
\end{equation}

Considering relation (\ref{eq:relation1}), we can multiply Eq.\
(\ref{eq:dirac}) by $\alpha^0$ on the left to get an alternate
form of Dirac equation
\begin{equation}
    \label{eq:altdirac}
    \frac{\mathrm{i}}{m} {\gamma_\kappa}^{\lambda \mu}
     \partial_\mu \psi_\lambda
    = \psi_\kappa,
\end{equation}
with ${\gamma_\kappa}^{\lambda \mu}$ defined by
\begin{equation}
    \label{eq:gammamatrix}
    {\gamma_\kappa}^{\lambda 0} = {\alpha_\kappa}^{\lambda 0};~~~~
    {\gamma_\kappa}^{\lambda i} = {\alpha_\beta}^{\lambda 0}
    {\alpha_\kappa}^{\beta i}.
\end{equation}
The $\gamma$ matrices obey the following relations
\begin{equation}
    \label{eq:gammarelations}
    {\gamma_\lambda}^{\beta \mu} {\gamma_\beta}^{\kappa \nu}
    + {\gamma_\lambda}^{\beta \nu} {\gamma_\beta}^{\kappa \mu} = 2
    \delta_\lambda^\kappa
    \eta^{\mu \nu}.
\end{equation}

Application of the $\mathrm{i} {\gamma_\kappa}^{\lambda \mu}
\partial_\mu /m$ operator on the left to both sides of
Eq.\ (\ref{eq:altdirac}) results in
\begin{equation}
    \label{seconddirac}
    \frac{ {\gamma_\kappa}^{\beta \mu}
    {\gamma_\beta}^{\lambda \nu}}{m^2}\,
    \partial_{\mu \nu} \psi_\lambda
    = -\psi_\kappa.
\end{equation}
Considering Eq.\ (\ref{eq:gammarelations}) and noting that
$\partial_{\mu \nu}= \partial_{\nu \mu}$ the equation can be
simplified to
\begin{equation}
    \label{eq:dirackg}
    \frac{ \delta_\kappa^\lambda
    \eta^{\mu \nu}}{m^2}\, \partial_{\mu \nu} \psi_\lambda
    = -\psi_\kappa;
\end{equation}
notice the striking similarity with Eq.\ (\ref{eq:founding}); the
factor $\delta_\kappa^\lambda$ in the first member serves only to
change the subscript of $\psi_\lambda$ and will be removed. The
source equation is then verified by all the elements $\psi_\kappa$
\begin{equation}
    \label{eq:source}
    g^{\mu
    \nu}\nabla_{\mu \nu} \psi_\kappa  = -\psi_\kappa;
\end{equation}
where $g^{\mu \nu} = \eta^{\mu \nu}/m^2$.

In order to accommodate gravitation we define the matrices
\begin{equation}
    \label{eq:smatrix}
    {s_\kappa}^{\lambda 0} = \frac{{\gamma_\kappa}^{\lambda 0}}{m n};~~~~
    {s_\kappa}^{\lambda i} = \frac{{\gamma_\kappa}^{\beta i}}{m};
\end{equation}
and modify Eq.\ (\ref{eq:altdirac}) to
\begin{equation}
    \label{eq:sdirac}
    \mathrm{i}{s_\kappa}^{\lambda \mu}
    \partial_\mu  \psi_\lambda
    = \psi_\kappa.
\end{equation}
The space metric is generalized to the 4-rank tensor
\begin{equation}
    \label{eq:smetric}
    {g_\kappa}^{\lambda \mu \nu} = {s_\kappa}^{\alpha \mu}
    {s_\alpha}^{\lambda \nu}.
\end{equation}
Finally, Rieman's condition $\nabla_\sigma g^{\mu \nu}=0$ is
replaced by the more general one
\begin{equation}
    \label{eq:genrieman}
    \nabla_\sigma {s_\kappa}^{\lambda \mu \nu}=0.
\end{equation}

Due to cancellation between crossed terms it will normally be
possible to use the simplified metric ${g_\kappa}^{\lambda \mu
\nu}=\mathrm{diag}(\mathrm{I}/n^2,-\mathrm{I},-\mathrm{I},-\mathrm{I})/m^2$
and even forget the $\kappa$ and $\lambda$ indices, using a
normal 2-rank tensor as metric.
\section{The point of view of 4-dimensional optics}
In Refs.\ \cite{Almeida02:1, Almeida02:2}, 4DO is developed in
4-space with signature $(+,+,+,+)$ and $\mathrm{d}t$ is defined
as being the line element in that space. Here we take a more
fundamental standing, considering the source equation
(\ref{eq:founding}) in the case where the metric defines a
pseudo-Euclidean space, i.e. with Euclidean tangent space; since
at this stage we are dealing with geometrical concepts, deprived
of any physical interpretations, we are allowed to try any metric
we decide to choose and any space signature we like. Later we
will have to check if the geometrical concepts can be given
physical interpretations, similarly to what has been done in
section \ref{interpretation}.

Let us assume a metric of the form
\begin{equation}
    \label{eq:4dometric}
    g^{\mu \nu} = \frac{n^2}{m^2}\, \delta^{\mu \nu};
\end{equation}
and insert this into Eq.\ (\ref{eq:founding}). As before assume
that $m$ is a constant and $n$ is a slowly varying function of
coordinates 1, 2 and 3; after simplification it is
\begin{equation}
    \label{eq:4dosimp}
    \frac{n^2}{m^2}\, \delta^{\mu \nu}
     \partial_{\mu \nu} \psi = -\psi.
\end{equation}
Now we can try a solution $\psi = \exp(\mathrm{i}p_\alpha
x^\alpha)$, resulting in
\begin{equation}
    \label{eq:4dosol}
    \frac{n^2}{m^2}\, \delta^{\mu \nu} p_\mu p_\nu = 1;
\end{equation}
from which we can write
\begin{equation}
    \label{eq:4domomenta}
    p_\mu = \frac{m^2}{n^2}\, \delta_{\mu \nu}\frac{\mathrm{d}
    x^\nu}{\mathrm{d} t},
\end{equation}
with the line element, designated by $\mathrm{d}t$, defined by
\begin{equation}
    \label{eq:dt2}
    \left(\mathrm{d}t \right)^2
    = \frac{m^2}{n^2}\, \delta_{\mu \nu} \mathrm{d}x^\mu
    \mathrm{d}x^\nu.
\end{equation}

Making a change of variables defined by Eqs.\ (\ref{eq:r},
\ref{eq:theta}, \ref{eq:phi}) together with
\begin{equation}
    \label{eq:tau}
    \tau = m x^0,
\end{equation}
proceeding as before, we arrive at the following Euler-Lagrange
equations:
\begin{gather}
    \label{eq:tau4do}
    \frac{1}{n^2} \frac{\mathrm{d}\tau}{\mathrm{d} t}
    = \frac{1}{\gamma}  ~~
      \rightarrow ~~ \mathrm{constant};\\
    \label{eq:phi4do}
    \frac{r^2}{n^2} \frac{\mathrm{d}\varphi}{\mathrm{d} t} =  J   ~~
      \rightarrow ~~ \mathrm{constant};\\
    \label{eq:rad4do}
    \frac{\mathrm{d}}{\mathrm{d} t}\left(\frac{1}{n^2}
    \frac{\mathrm{d} r}{\mathrm{d}t}\right)=
    -n\frac{\mathrm{d}n}{\mathrm{d} r}\
    + \frac{n^2 J^2}{r^3}.
\end{gather}
Equations (\ref{eq:tau4do}) to (\ref{eq:rad4do}) are equivalent
to Eqs.\ (\ref{eq:taual}) to (\ref{eq:radial}); considering $n$
defined by Eq.\ (\ref{eq:gravpot}) we obtain Eq.\
(\ref{eq:radial2}) as before.

It is important to verify that Poisson equation is not violated
if the covariant Laplacian is used in place of the usual one. The
covariant Laplacian of a scalar function $f$ is evaluated as
\cite{Nakahara90}
\begin{equation}
    \label{eq:laplacian}
    g^{\mu \nu} \nabla_{\mu \nu} f = - \frac{1}{\sqrt{g}}\,
    \partial_\mu \left( \sqrt{g} g^{\mu \nu} \partial_\nu f
    \right).
\end{equation}
Using $g^{\mu \nu}$ defined by Eq.\ (\ref{eq:4dometric}) and
evaluating the covariant Laplacian of $n$ given by Eq.\
(\ref{eq:gravpot}) one gets
\begin{equation}
    \label{eq:poisson}
    g^{\mu \nu} \nabla_{\mu \nu} n^2 = 0,
\end{equation}
as expected.

Neither Klein-Gordon or Dirac equation is compatible with 4DO,
basically due to the different space signature. With respect to
Klein-Gordon equation, its 4DO counterpart is obviously obtained
from Eq.\ (\ref{eq:founding}) when $g^{\mu \nu} = \delta^{\mu
\nu}$. In order to obtain a 4DO equation which accounts for spin
we replace ${\gamma_\kappa}^{\lambda \mu}$ by
${\alpha_\kappa}^{\lambda \mu}$ in  Dirac equation
(\ref{eq:altdirac}) such that
\begin{equation}
    \label{eq:4dodirac}
    \frac{\mathrm{i}}{m}\, {\alpha_\kappa}^{\lambda \mu}
     \partial_\mu \psi_\lambda
    = \psi_\kappa.
\end{equation}
A procedure similar to what is used in textbooks to derive Dirac
equation's solutions for a particle at rest \cite{Cottingham98}
would show that Eq,\ (\ref{eq:4dodirac}) has solutions of the form
\begin{equation}
    \label{eq:diracsol}
    \psi_\kappa = e^{\pm \mathrm{i}m x^0} u_\kappa,
\end{equation}
when $\partial_i \psi_\kappa =0$, corresponding to a particle at
rest. For the positive exponent, $u_\kappa$ has one of the two
forms
\begin{equation}
    \label{eq:uforms}
    u_\kappa = (1,0,1,0),~~~~ u_\kappa = (0,1,0,1),
\end{equation}
corresponding to the two spin possibilities.

Equation (\ref{eq:sdirac}) can be applied in 4DO if the matrices
${s_\kappa}^{\lambda \mu}$ are redefined as
\begin{equation}
    \label{eq:4dos}
    {s_\kappa}^{\lambda \mu}= \frac{n {\alpha_\kappa}^{\lambda
    \mu}}{m}.
\end{equation}
Cancellation between crossed terms will normally allow the use of
the 2-rank metric defined by Eq.\ (\ref{eq:4dometric}).
\section{Electrodynamics}
4-dimensional optics handles electrodynamics in a way that
differs substantially from the relativistic approach, the latter
being synthesized in the substitution $p^\mu \rightarrow p^\mu - q
A^\mu$, with $p^\mu$ the contravariant momentum, $q$ electric
charge and $A^\mu$ a vector potential. We shall demonstrate that
Eq.\ (\ref{eq:sdirac}) can account for electromagnetic
interaction in 4DO and will not attempt a similar approach in
relativistic space-time, which we believe is not possible. We
propose that the ${s_\kappa}^{\lambda \mu}$ matrices should be
defined as
\begin{equation}
    \label{eq:elmags}
    {s_\kappa}^{\lambda 0} = \frac{{\alpha_\kappa}^{\lambda \mu}
    V_\mu}{m},~~~~
    {s_\kappa}^{\lambda i} = \frac{{\alpha_\kappa}^{\lambda
    i}}{m}.
\end{equation}
Application of Eq.\ (\ref{eq:smetric}) and cancellation leads to
the following metric
\begin{equation}
    \label{emmetric}
    g^{\mu \nu} = \frac{1}{m^2}\left(\begin{array}{cccc}
      \delta^{\kappa \lambda} V_\kappa V_\lambda
      &  V_1 & V_2
        & V_3 \\
      V_1 & 1 & \cdot & \cdot \\
      V_2 & \cdot & 1 & \cdot \\
      V_3 & \cdot & \cdot & 1 \
    \end{array} \right).
\end{equation}

The lower subscript metric $g_{\mu \nu}$ is obtained, as usual,
by calculating the inverse of $g^{\mu \nu}$.
\begin{widetext}
\begin{equation}
    \label{eq:eminverse}
    g_{\mu \nu} = \frac{m^2}{(V_0)^2}\left(\begin{array}{cccc}
      1 & -V_1
      & -V_2 & -V_3 \\
      -V_1 & (V_0)^2 +(V_1)^2
      & V_1 V_2 & V_1 V_3 \\
      -V_2 & V_1 V_2
      & (V_0)^2 + (V_2)^2 & V_2 V_3 \\
      -V_3 & V_1 V_3
      & V_2 V_3 & (V_0)^2 + (V_3)^2 \
    \end{array}\right).
\end{equation}
\end{widetext}

Let us consider the situation where $V_i = 0$ and write the line
element
\begin{equation}
    \label{eq:elelement0}
    \left(\mathrm{d}s\right)^2=\frac{m^2}{(V_0)^2} \left(\mathrm{d} x^0
     \right)^2 + m^2 \delta_{ij} \mathrm{d}
    x^i\mathrm{d}x^j;
\end{equation}
next we change coordinates ($x^{\bar{\mu}} = m x^\mu$); following
the usual procedure we write the Lagrangian
\begin{equation}
    \label{eq:eleclag}
    2L = \frac{1}{(V_0)^2} \left(\frac{\mathrm{d} x^{\bar{0}}}
    {\mathrm{d} t} \right)^2 +  \delta_{\bar{i}\bar{j}} \frac{\mathrm{d}
    x^{\bar{i}}}{\mathrm{d}t}\frac{\mathrm{d}x^{\bar{j}}}{\mathrm{d}t}.
\end{equation}
The equations follow directly
\begin{gather}
    \label{eq:elec0}
    p_{\bar{0}} = \frac {1}{(V_0)^2} \frac{\mathrm{d}x^{\bar{0}}}{\mathrm{d} t}
    = \frac{1}{\gamma},~~\rightarrow~~\mathrm{constant},\\
    \label{eq:eleci}
    \frac{\mathrm{d}^2 x^{\bar{i}}}{\mathrm{d}t^2} =
    -\frac{V_0}{\gamma^2}\, \partial_{\bar{i}} V_0.
\end{gather}

Equation (\ref{eq:eleci}) represents a particle's dynamics under
an electric field if
\begin{equation}
    \label{eq:elpot}
    V_0 = \mathrm{e}^{- q A_0},
\end{equation}
with $q$ the electric charge and $A_0$ the electric potential.
Inserting in (\ref{eq:eleci})
\begin{equation}
    \label{eq:eleci2}
    \frac{\mathrm{d}^2 x^{\bar{i}}}{\mathrm{d}t^2} =
    \frac{q}{\gamma^2}\, \mathrm{e}^{-2 q A_0} \partial_{\bar{i}} A_0.
\end{equation}

The fact that the electric potential appears as exponent of the
metric coefficient $V_0$ is seen as a consequence of it too being
the manifestation of evanescence from a 4-dimensional waveguide
\cite{Almeida02:1}; if we compare Eq.\ (\ref{eq:eleci}) to the
radial equation obtained for gravity (\ref{eq:rad4do}) we can
immediately find obvious similarities. The electric potential has
a typical $1/r$ rule and consequently we will consider $V_0 =1$
far away from any electric charges.

As a second example we consider now the situation where
$\partial_{\bar{\mu}} V_\nu = 0$ except for $\partial_{\bar{1}}
V_2 = q B_3$; for simplicity we make $V_0 =1$, $V_1 = V_3 = 0$ and
$V_2 = x^{\bar{1}} q B_3$; $B_3$ represents the magnetic field
aligned along $x^3$. The Langrangian is now defined by
\begin{equation}
\begin{split}
    \label{eq:maglag}
    2L =\, & \delta_{\bar{\mu} \bar{\nu}}
    \frac{\mathrm{d}x^{\bar{\mu}}}{\mathrm{d}t}
    \frac{\mathrm{d}x^{\bar{\nu}}}{\mathrm{d}t}\\
    & -2q x^{\bar{1}} B_3 \frac{\mathrm{d}x^{\bar{0}}}{\mathrm{d}t}
    \frac{\mathrm{d}x^{\bar{2}}}{\mathrm{d}t}
    + \left(q x^{\bar{1}} B_3 \frac{\mathrm{d}x^{\bar{2}}}{\mathrm{d}t}
    \right)^2.
\end{split}
\end{equation}

Some straightforward calculations lead to the conjugate momenta
\begin{align}
    \label{eq:magmom0}
    & p_{\bar{0}} = \frac{\mathrm{d}x^{\bar{0}}}{\mathrm{d}t}
    -q x^{\bar{1}}B_3 \frac{\mathrm{d}x^{\bar{2}}}{\mathrm{d}t},\\
    \label{eq:magmom1}
    & p_{\bar{1}}= \frac{\mathrm{d}x^{\bar{1}}}{\mathrm{d}t},\\
    \label{eq:magmom2}
    & p_{\bar{2}}= \frac{\mathrm{d}x^{\bar{2}}}{\mathrm{d}t}
    -q x^{\bar{1}} B_3 \frac{\mathrm{d}x^{\bar{0}}}{\mathrm{d}t}
    + \left(q x^{\bar{1}} B_3 \right)^2
    \frac{\mathrm{d}x^{\bar{2}}}{\mathrm{d}t},\\
    \label{eq:magmom3}
    & p_{\bar{3}}= \frac{\mathrm{d}x^{\bar{3}}}{\mathrm{d}t}.
\end{align}
Since the Lagrangian is independent from $x^{\bar{3}}$, from Eq.\
(\ref{eq:magmom3}) one gets
\begin{equation}
    \label{eq:mag3}
    \frac{\mathrm{d}x^{\bar{3}}}{\mathrm{d}t} = \mathrm{constant};
\end{equation}
then, independence of the Lagrangian from $x^{\bar{0}}$ and Eq.\
(\ref{eq:magmom0}) give
\begin{equation}
    \label{eq:mag0}
    \frac{\mathrm{d}x^{\bar{0}}}{\mathrm{d}t} = \frac{1}{\gamma} +
    q x^{\bar{1}} B_ 3 \frac{\mathrm{d}x^{\bar{2}}}{\mathrm{d}t},
\end{equation}
with $\gamma$ constant.

Inserting Eq.\ (\ref{eq:mag0}) into Eq.\ (\ref{eq:magmom2}) and
deriving with respect to $t$
\begin{equation}
    \label{eq:mag2}
    \frac{\mathrm{d}^2 x^{\bar{2}}}{\mathrm{d}t^2} =
    \frac{q B_3}{\gamma} \frac{\mathrm{d}x^{\bar{1}}}{\mathrm{d}t}.
\end{equation}
Finally, from Eq.\ (\ref{eq:magmom1}), deriving the Lagrangian
with respect to $x^{\bar{1}}$ and making the substitutions
\begin{equation}
    \label{eq:mag1}
    \frac{\mathrm{d}^2 x^{\bar{1}}}{\mathrm{d}t^2} =
    \frac{q B_3}{\gamma} \frac{\mathrm{d}x^{\bar{2}}}{\mathrm{d}t}.
\end{equation}
Equations (\ref{eq:mag2}) and (\ref{eq:mag1}) represent the
Lorentz force exerted by the magnetic field $B_3$ over a particle
of electric charge $q$ and moving with velocity
$\mathrm{d}x^{\bar{i}}/\mathrm{d}t$.

The two examples show that the definition in Eq.\ (\ref{eq:4dos})
generates a metric whose geodesics are the worldlines of
particles under electromagnetic field. Electrodynamics is
associated to space curvature, in much the same way as gravity. In
general one can define an electromagnetic tensor akin to its
relativistic counterpart
\begin{equation}
    \label{eq:emtensor}
    F_{\mu \nu} = \frac{\partial_\mu V_\nu - \partial_\nu
    V_\mu}{q} = \frac{\nabla_\mu V_\nu - \nabla_\nu
    V_\mu}{q}.
\end{equation}
The exponential factor $e^{q A_0}$ will normally be too near unity
to be of any practical consequence in everyday situations. Note
that $F_{\mu \nu}$ has been defined in observer space; the
transition between dynamic and observer spaces occurs when we
divide by $q$. Similarly to what we commented about inertial
mass, mapping between observer and dynamic spaces implies the
presence of an electric charge.
\section{Conclusion and further work}
The dynamic space concept, introduced in this paper allows
quantum mechanics and gravitation to be reconciled under the
umbrella of the same equations; this was first checked with
Klein-Gordon equation, through the definition of a gravitational
law, already presented in previous work. It was shown that
Klein-Gordon equation presumes dynamic space, while classical
mechanics is normally set in observer space. The compatibility of
Dirac equation with dynamic space and the new gravitational law
was also established.

4-dimensional optics was shown to be an alternative dynamic space
where both quantum and classical mechanics could be set with
equivalent predictions to those made in the more usual space-time
for all but extreme situations, implying a modification of Dirac
equation; in this alternative space the new gravitational law was
shown to obey Poisson equation. The electrodynamics of a charged
particle was analyzed in 4-dimensional optics and was shown to
verify the modified Dirac equation but also to be amenable to
space curvature, similarly to gravity.

The work presented in this paper is largely incomplete. Several
years and great effort are needed to extrapolate the conclusions
to cases of mass and charge distributions, to draw cosmological
consequences and to work out the compatibility with the standard
model of particle physics. In this respect we believe that
elementary particles must be seen as dynamic spaces on their own
and the space metrics generated by the solutions of the
fundamental equations must then be converted into observer space
through suitable coordinate changes; suggestions in this respect
have already been made in Ref.\ \cite{Almeida02:1} and prior to
that work mass has been shown to emerge from 4-dimensional
guidance of a particle's wavefunction \cite{Almeida01:5}. The
author believes that all physics must be revised under the new
perspectives and hopes to contribute to this revision in future
papers but this is task too big for a single person to tackle;
the present work, albeit incomplete, should set the bases for
anybody who feels motivated to join in the effort.
%

  \bibliography{abrev,aberrations,assistentes}   

\end{document}